\documentclass[letterpaper,twocolumn,10pt]{article}
\usepackage{style/usenix}

\usepackage{algorithm}
\usepackage{amsmath}
\usepackage{amsthm}
\usepackage{amsfonts}
\usepackage{amssymb}

\usepackage{color}
\usepackage{enumitem}
\usepackage{fancyhdr}
\usepackage{float}
\usepackage{graphicx}
\usepackage{hyperref}

\usepackage{listings}

\usepackage[binary-units]{siunitx}
\usepackage{url}
\usepackage{xspace}
\usepackage{hyphenat}

\lstdefinelanguage{javascript}{
  keywords={typeof, new, true, false, catch, function, return, null, catch, switch, var, if, in, while, do, else, case, break},
  keywordstyle=\color{blue}\bfseries,
  ndkeywords={class, export, boolean, throw, implements, import, this},
  ndkeywordstyle=\color{black}\bfseries,
  identifierstyle=\color{black},
  sensitive=false,
  comment=[l]{//},
  morecomment=[s]{/*}{*/},
  commentstyle=\color{purple}\ttfamily,
  stringstyle=\color{red}\ttfamily,
  morestring=[b]',
  morestring=[b]"
}

\newcommand\myurl[2]{\url{#1}}

\newfloat{lstfloat}{htbp}{lop}
\floatname{lstfloat}{Listing}

\newenvironment{description*}%
  {\begin{description}%
    \setlength{\itemsep}{0pt}%
    \setlength{\parskip}{0pt}}%
  {\end{description}}
  
\newcommand{\anonword}[2]{%
\expandafter\newcommand\csname #1\endcsname{#2\xspace}%
}%

\begin{document}

\title{On-demand Container Loading in AWS Lambda}

\newcommand{\awsauthor}[2]{
{\rm #1}\\
Amazon Web Services
}
\makeatletter
\iffalse
\anonword{enclaves}{Anoncloud TEE}
\anonword{AWS}{Anoncloud}
\anonword{Amazon}{Anoncloud}
\anonword{SThree}{AnonStore}
\anonword{AWSLambda}{AnonFaaS}
\else
\anonword{enclaves}{AWS Nitro Enclaves}
\anonword{AWS}{AWS}
\anonword{Amazon}{Amazon}
\anonword{SThree}{S3}
\anonword{AWSLambda}{Lambda}
\fi

\iftrue
\author{
\awsauthor{Marc Brooker}{mbrooker@amazon.com}
\and
\awsauthor{Mike Danilov}{mikhaild@amazon.com}
\and
\awsauthor{Chris Greenwood}{greenwd@amazon.com}
\and
\awsauthor{Phil Piwonka}{piwonka@amazon.com}
}
\else

\author{
\awsauthor{Anonymous Authors}{@amazon.com}
\awsauthor{Eric Heinz}{eheinz@amazon.com}

}
\fi

\makeatother

\markboth{DRAFT. Amazon Confidential. Not for distribution or publication.}%
{DRAFT. Amazon Confidential. Not for distribution or publication.}

\maketitle

\section*{Abstract}
\AWS \AWSLambda is a serverless event-driven compute service, part of a category of cloud compute offerings sometimes called Function-as-a-service (FaaS). When we first released \AWS \AWSLambda, functions were limited to 250MB of code and dependencies, packaged as a simple compressed archive. In 2020, we released support for deploying container images as large as 10GiB as \AWSLambda functions, allowing customers to bring much larger code bases and sets of dependencies to \AWSLambda. Supporting larger packages, while still meeting \AWSLambda's goals of rapid scale (adding up to 15,000 new containers per second for a single customer, and much more in aggregate), high request rate (millions of requests per second), high scale (millions of unique workloads), and low start-up times (as low as 50ms) presented a significant challenge.

We describe the storage and caching system we built, optimized for delivering container images on-demand, and our experiences designing, building, and operating it at scale. We focus on challenges around security, efficiency, latency, and cost, and how we addressed these challenges in a system that combines caching, deduplication, convergent encryption, erasure coding, and block-level demand loading.

Since building this system, it has reliably processed hundreds of trillions of \AWSLambda invocations for over a million AWS customers, and has shown excellent resilience to load and infrastructure failures.
\section{Introduction}

\AWS \AWSLambda is a serverless event-driven compute service, part of a category of cloud compute offerings sometimes called Function-as-a-service (FaaS). First launched in 2015, today \AWS \AWSLambda functions run millions of times per second over millions of unique customer workloads. One factor that attracts customers to \AWSLambda is its ability to scale up to handle increased load, typically in less than one second (and often as quickly as 50ms).
This scale-up time, which customers have come to refer to as \emph{cold-start time}, is one of the most important metrics that determine the customer experience in FaaS systems. When we launched \AWS \AWSLambda, we recognized that reducing data movement during these cold starts was critical. Customers deployed functions to \AWSLambda in compressed archives (\emph{.zip} files), which were unpacked as each function instance was provisioned.
As \AWSLambda evolved, and customers increasingly looked to deploy more complex applications, there was significant demand for larger deployments, and the ability to use container tooling (such as \emph{Docker}) to create and manage these deployment images. Customers also wanted \AWSLambda to support these images without compromising on cold-start performance.

Adding container support to \AWS \AWSLambda without regressing on cold-start time presented a significant technical challenge for our team. The core challenge is simply one of data movement. Today,  \AWSLambda can start up to 15,000 containers a second~\cite{Ionescu2022} for production workloads, and we expect to scale further for future workloads. Simply moving and unpacking a 10GiB image for each of these 15,000 containers would require 150Pb/s of network bandwidth. To achieve scalability and cold-start latency goals, we needed to take advantage of three factors which simplify this problem:

\begin{description}
\item[Cacheability] While  \AWSLambda serves hundreds of thousands of unique workloads, large scale-up spikes tend to be driven by a smaller number of images, suggesting that the workload is highly cacheable.
\item[Commonality] Many popular images are based on common base layers (such as our own \AWS base layers, or open source offerings like Alpine). Caching and deduplicating these common base layers reduce data movement for all containers that build on them.
\item[Sparsity] Most container images contain a lot of files, and file contents, that applications don't need at startup (or potentially never need). Harter et al~\cite{Harter2016} found that on average only 6.4\% of container data is needed at startup.
\end{description}

Our solution combines caching, deduplication, erasure coding, and sparse loading to take advantage of our needs. Without adding any customer visible complexity (they simply upload a container image to a convenient repository), we were able to achieve our scale and cold-start latency goals, while having significant headroom for future scaling.

In this section, we present the existing architecture of \AWS \AWSLambda, and the overall architecture of our system. Section \ref{sec:blocklevel} presents the low-level implementation of our sparse loading solution. The cache architecture, and use of erasure coding to improve scalability and tail latency is presented in Section \ref{sec:cache}. Section \ref{sec:dedupe} presents our convergent encryption-based secure deduplication architecture. Finally Section \ref{sec:related} compares our solution to other approaches from academia and industry.

\subsection{Existing Architecture Overview}
\begin{figure}[!t]
\centering
\includegraphics[width=0.6\columnwidth]{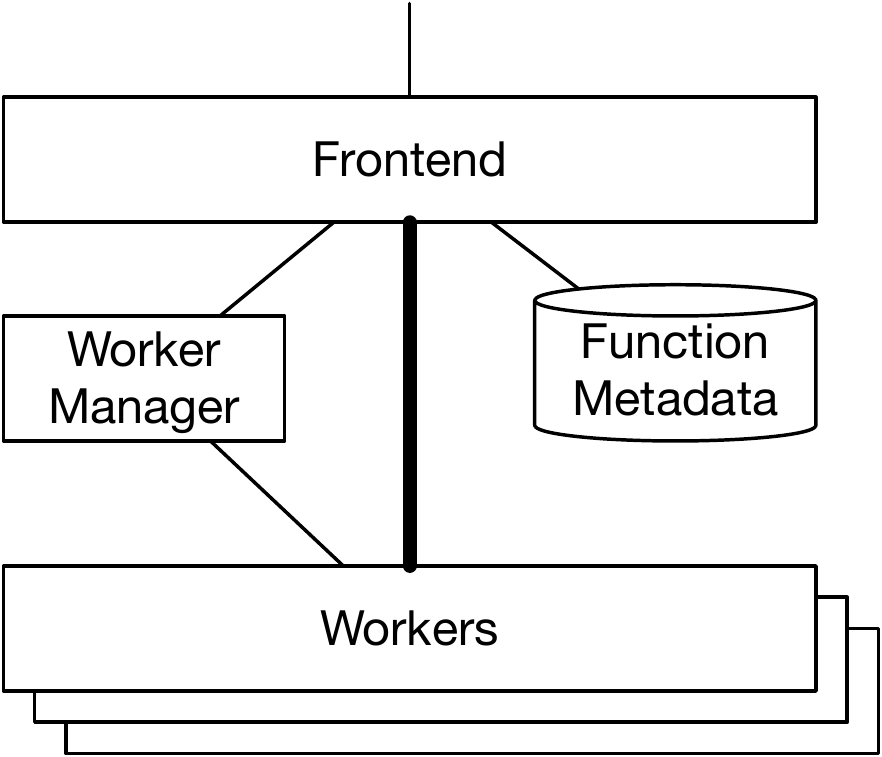}
\caption{Architecture of the \AWS \AWSLambda invoke path}
\label{fig:architecture}
\end{figure}

To reduce risk and optimize time-to-market, we wanted to introduce these new capabilities to \AWSLambda with the minimum amount of change to the existing architecture, as shown in Figure~\ref{fig:architecture}. Requests to execute a certain function (we call these \emph{invokes}) arrive via a load-balanced stateless frontend service. This service loads the metadata associated with the request, performs authentication and authorization, and then sends a request to the \emph{Worker Manager}, requesting capacity. \emph{Worker Manager} is a stateful, sticky, load balancer. For every unique function in the system, it keeps track of what capacity is available to run that function, where that capacity is in the fleet, and predicts when new capacity may be needed. If capacity is available, the Worker Manager instructs the frontend to forward the request payload to a Worker, where the function is executed. If no capacity is available, the Worker Manager identifies a Worker with available CPU and RAM, and sends a request to start a sandbox for the relevant function. Once this is complete, the frontend is notified and the function is executed.

\begin{figure}[!t]
\centering
\includegraphics[width=0.8\columnwidth]{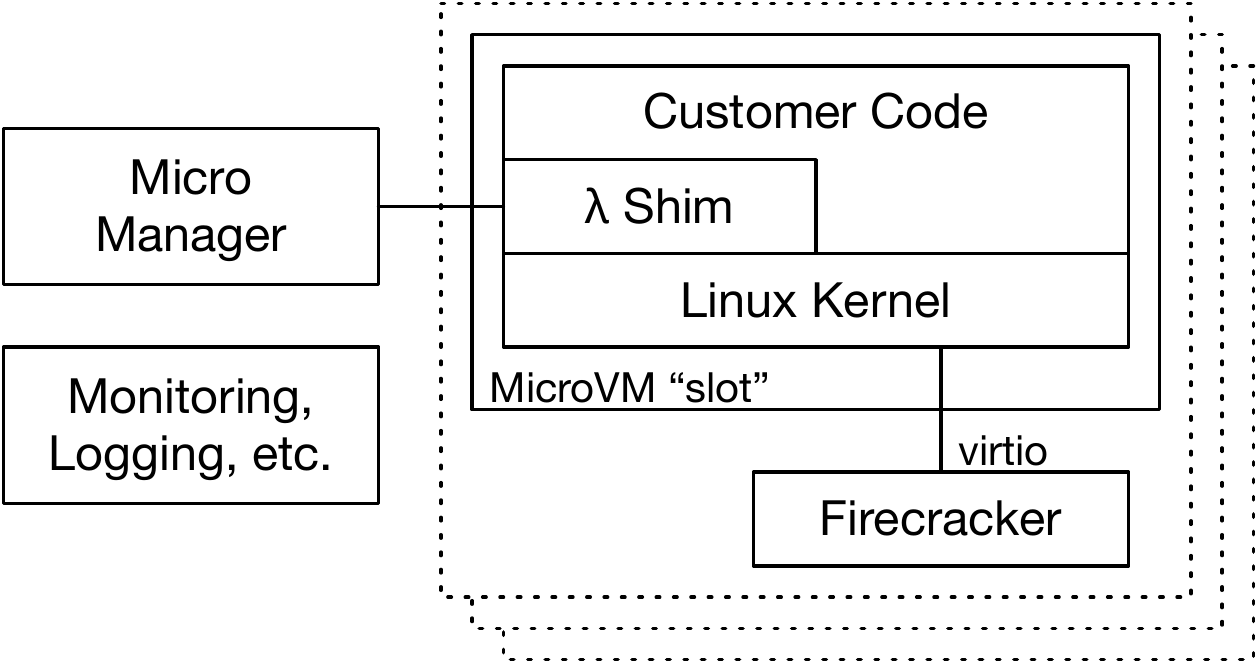}
\caption{Architecture of the \AWS \AWSLambda worker}
\label{fig:worker_arch}
\end{figure}

Each \AWSLambda worker, as shown in Figure \ref{fig:worker_arch}, includes a small controller process, the Micro Manager, some additional agents for logging and monitoring, and a large number of MicroVMs. Each MicroVM, based on our Firecracker~\cite{agache2020} hypervisor, contains the code for a single \AWSLambda function for a single customer. Inside the MicroVM is a minimized Linux guest kernel, a small shim that provides \AWSLambda's programming model, any provided runtime (e.g. the JVM for Java or CoreCLR for .NET), and the customer's code and libraries. As described in our Firecracker paper~\cite{agache2020}, the key concern here is security: customer code and data is not trusted, and the only communication between the workload inside the MicroVM and the shared worker components is over a simple, well tested, and formally verified implementation of \emph{virtio}~\cite{Russell2008, virtio2016} (specifically \emph{virtio-net} and \emph{virtio-blk}).

In the first generation architecture (before this work), when a new MicroVM is created with new capacity for a particular function, the Worker downloads the function image (a \emph{.zip} file up to 250MiB in size) from \Amazon \SThree, and unpacks it into the MicroVM guest's filesystem. This model is simple, and works well for small images, but requires the full archive to be downloaded and unpacked before the new MicroVM can do any work. To support larger images, we wanted to avoid this blocking download, and avoid the storage cost of unpacking the entire archive if only part of it is used.

\section{Block-Level Loading}\label{sec:blocklevel}

\begin{figure}[!t]
\centering
\includegraphics[width=0.6\columnwidth]{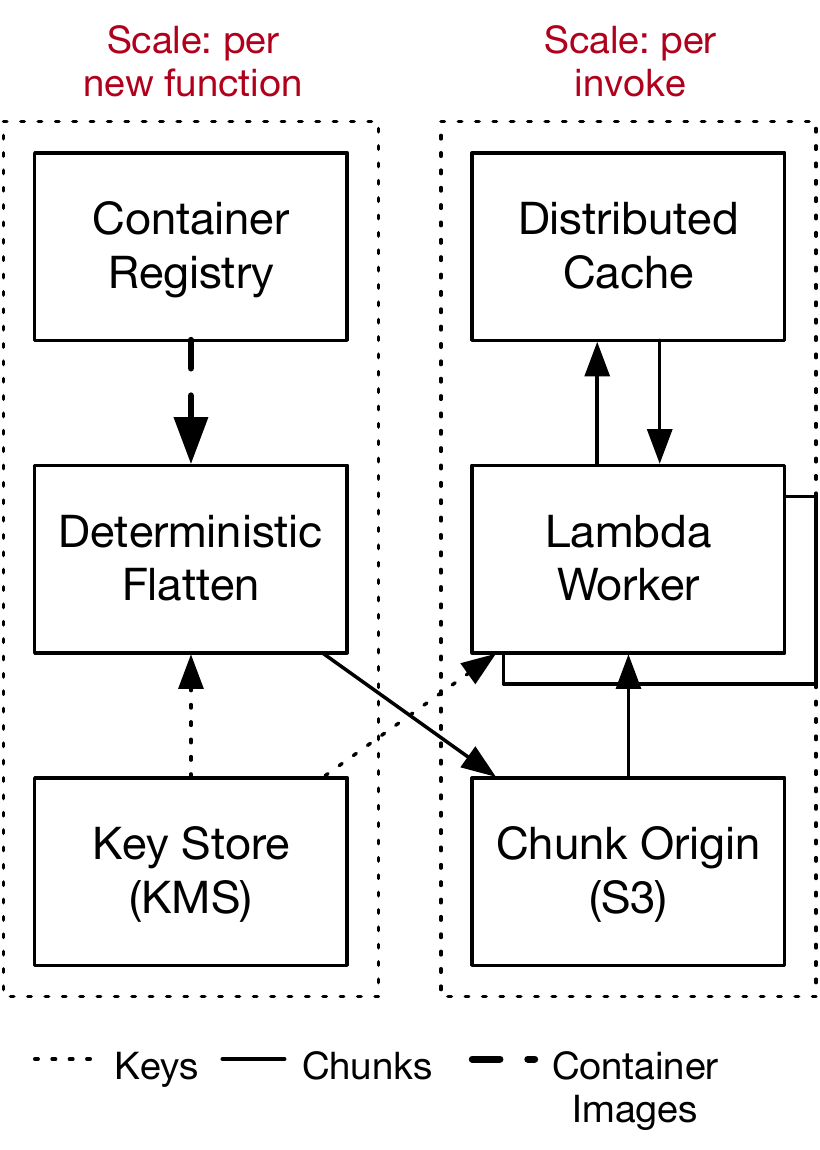}
\caption{High-level system architecture.}
\label{fig:high_level_arch}
\end{figure}

To take advantage of the \emph{sparsity} property of containers, we needed to allow the system to load (and store) only the data the application needs, ideally at the time it needs it. Approaches like Slacker~\cite{Harter2016} and Starlight~\cite{Chen2022} have approached this problem at the filesystem level - a natural fit for containers, which are built as an overlayed stack of file-level archives. This approach isn't the right one for our environment. We believed that the inherent complexity of filesystems, and additional complexity of overlaying multiple filesystems, would unacceptably increase the attack surface of the shared components in \AWSLambda. Instead, we decided to keep the block-level \emph{virtio-blk} interface between the MicroVM guest and the hypervisor, perform all filesystem operations inside the guest. This requires performing sparse loading at the block, rather than file, level.

\begin{figure}[!t]
\centering
\includegraphics[width=0.9\columnwidth]{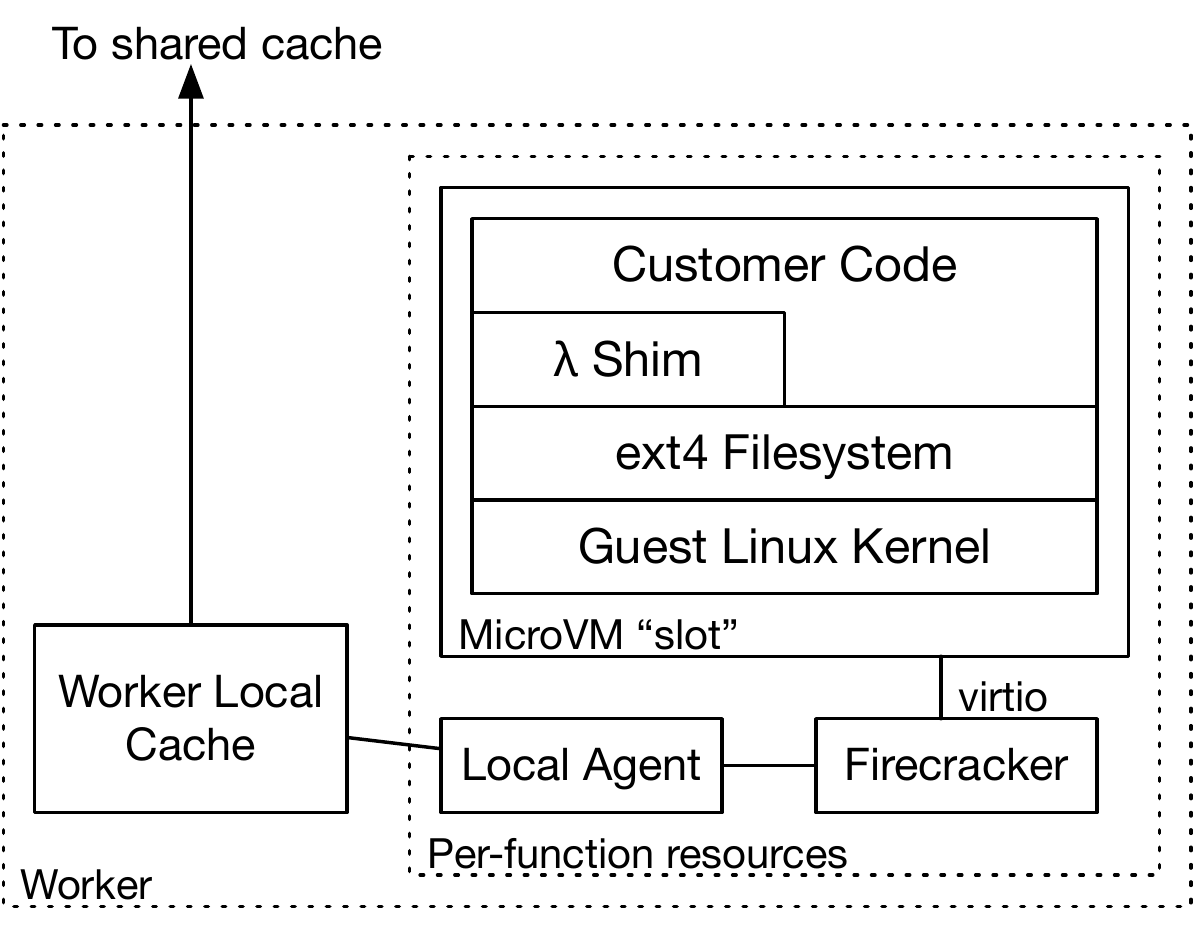}
\caption{Lambda worker with per-worker, per-customer, and in-guest components}
\label{fig:on_worker_arch}
\end{figure}

Figure \ref{fig:high_level_arch} shows our high-level architecture, showing the \AWSLambda worker (shown in detail in Figure \ref{fig:on_worker_arch}) where customer's code is run, container registry which contains the primary copy of customer's container images, and the chunk creation and caching infrastructure.

Our first step in supporting block-level loading is to collapse the container image into a block device image. As described in the OCI image specification~\cite{OCISpec}, a container image is a stack of tarball \emph{layers}. In the typical container stack, these layers are overlayed at at runtime using \emph{overlayfs}. In our implementation, we perform this overlaying operation at the time the function is initially created, following a deterministic flattening process which applies each tarball in order to create a single \emph{ext4} filesystem. Function creation is a low-rate control-plane process, that is typically only triggered by customers when they make changes to their code, configuration, or architecture. Even the most aggressive adoptees of continuous integration only make these changes on the order of minutes, while function invocation can happen up to millions of times a second. 

The flattening process is designed so that blocks of the filesystem that contain unchanged files will be identical, allowing for block-level deduplication of the flattened images between containers that share common base layers. We'll revisit this in Section \ref{sec:dedupe}, but the high-level reason is that differences between functions (and even more so between versions of the same function) are typically much smaller than the functions themselves. The flattening process proceeds by unpacking each layer onto an \emph{ext4} filesystem, using a modified filesystem implementation that performs all operations deterministically. Most filesystem implementations take advantage of concurrency to improve performance, introducing non-determinism. Ours is serial, and deterministically chooses normally-variable parameters like modification times.

Following the flattening process, the flattened filesystem is broken up into fixed-size chunks, and those chunks are uploaded to the origin tier of a three-tiered cache for later use (we use \SThree as this origin tier). Chunks in the shared storage are named according the their content, ensuring that chunks with the same content have the same name and can be cached once. This scheme, described in detail in Section~\ref{sec:dedupe}, allows efficient deduplication of chunk content in storage and cache layers without requiring a central directory or index of chunks.

Each fixed-size chunk is 512KiB. Smaller chunks lead to better deduplication by minimizing false-sharing, and can accelerate loading for workloads with highly random access patterns. Larger chunks reduce metadata size, reduce the number of requests needed to load data (hence improving throughput), and provide natural read-ahead for sequential workloads. The optimal value will change over time as the system evolves, and we expect that future iterations of the system may choose a different chunk size as our understanding of how customers use the system evolves.

\subsection{Per-MicroVM Snapshot Loading}
Once chunks are created, the system needs to be able to access the data they require from the chunks that contain that data. As shown in Figure \ref{fig:on_worker_arch}, we added two new components to support this loading: 

\begin{itemize}
\item A per-function \emph{local agent} which presents a block device to the per-function Firecracker hypervisor (via FUSE), which is then forwarded using the existing virtio interface into the guest, where it is mounted by the guest kernel.
\item A per-worker \emph{local cache} which caches chunks of data that are frequently used on the worker, and interacts with the remote cache (see Section \ref{sec:cache} for details)
\end{itemize}

When a new \AWSLambda function is started on a worker, the Micro Manager creates a new \emph{local agent}, and a new Firecracker MicroVM which contains two \emph{virtio} block devices: a root device which is the same for all MicroVMs, and a block device backed by the FUSE filesystem exposed by the \emph{local agent}. The MicroVM boots, starts some supervisory components, and then starts executing the customer code in the container image. Each IO that this code performs (unless it can be served from the page cache kept by the guest kernel) turns into a \emph{virtio-blk} request, which is then processed by Firecracker, and handed off to the local agent.

The local agent handles reads by reading directly from the local cache, if the chunk that contains the requested offset is already present there. If not, the relevant chunk is fetched from the tiered cache, as described in Section \ref{sec:cache}. The local agent handles write by writing them to block overlay, backed by encrypted storage on the worker. A bitmap is maintained at page granularity, indicating whether data should be read from the overlay, or from the backing container image. The page granularity of the bitmap requires a read-modify-write for writes from the guest which don't cover an entire page.

This page-level copy-on-write approach allows the MicroVM guest to handle both reads and writes, while keeping the data in the local cache (and all other caching tiers) immutable, allowing it to be shared across multiple guests.

\section{Deduplication Without Trust}~\label{sec:dedupe}
Base container images, such as the official Docker \emph{alpine}, \emph{ubuntu}, and \emph{nodejs} are extremely widely used: each boasts over a billion aggregate downloads from the popular DockerHub container repository\footnote{statistics from https://hub.docker.com/, accessed July 2022}. Starting from one of these base images, and customizing it to the special needs of the application, is a common way to create new container images. When a popular base image is used, the deterministic flattening process described in Section~\ref{sec:blocklevel} produces unique chunks for the customized parts, and chunks for the common parts that are identical to those produced for other images with the same base. These shared chunks create a significant opportunity for deduplication: if only a single copy of these chunks is stored, less data movement is needed, less storage is consumed, and caches are more effective.

Approximately 80\% of newly uploaded Lambda functions result in zero unique chunks, and are just re-uploads of images that had been uploaded in the past. This appears to be primarily driven by automated testing and deployment (CI/CD) systems. Of the remaining 20\% of functions that create at least one unique chunk (and therefore aren't just trivial re-uploads), the mean upload contains 4.3\% unique chunks, and the median 2.5\% unique chunks. Trivial all-zero chunks are not included in these numbers: they are excluded entirely from images at creation time.

\begin{figure}[!t]
\centering
\includegraphics[width=0.8\columnwidth]{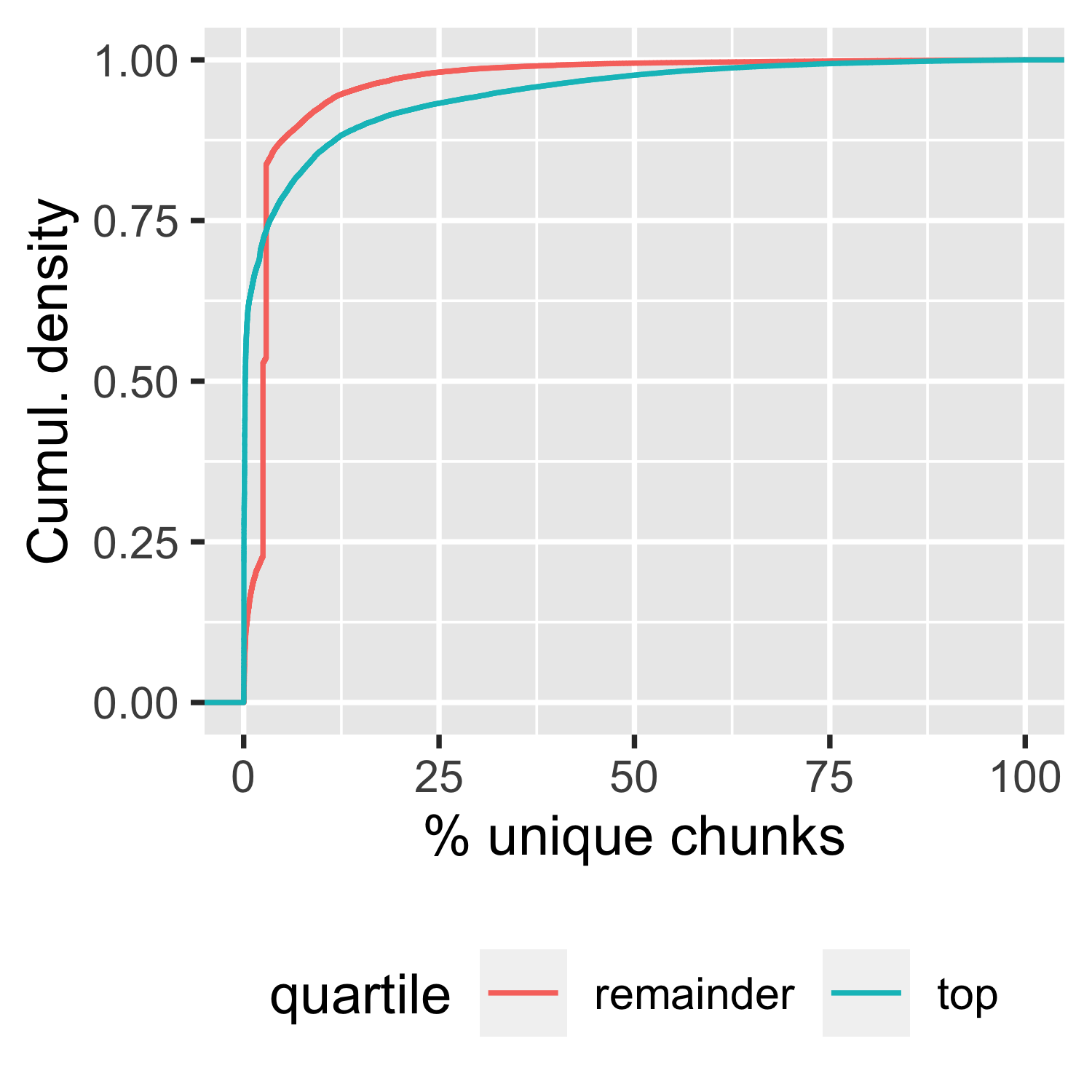}
\caption{Empirical CDF of deduplication effectiveness at chunk creation time, among functions that aren't trivial re-uploads.}
\label{fig:dedup_cdf}
\end{figure}

Figure \ref{fig:dedup_cdf} shows the distribution of deduplication effectiveness, for the top quartile (by image size) and remainder of the population. This breakdown shows that the majority of functions of all sizes are heavily deduped, and a significant tail where deduplication is not as effective. While large functions are still effectively deduplicated, they have a smaller tail of unique chunks. This data clearly suggests that deduplication is worth the complexity, reducing storage by as much as 23x, and improving effectiveness of the cache tiers (how much cache effectiveness is improved depends on the correlation between probability of deduplication and frequency of access). While the 80\% of functions with no unique chunks aren't statistically interesting, deduplicating these has a large practical benefit, including reducing storage costs by another 5x, and boosting cache effectiveness.

\subsection{Convergent Encryption}
Deduplication of plaintexts is relatively straightforward. Venti~\cite{Quinlan2002}, dating back to 2002, used a hash of block contents and a separate index to deduplicate blocks. Introducing encryption, however, significantly complicates deduplication. As Storer, et al~\cite{Storer2008} write:

\begin{quote}
Unfortunately, deduplication exploits identical content, while encryption attempts to make all content appear random; the same content encrypted with two different keys results in very different ciphertext. Thus, combining the space efficiency of deduplication with the secrecy aspects of encryption is problematic.
\end{quote}

One solution is to have a shared key, or keys, that can be used to decrypt shared blocks, but this either introduces single keys that can access a large number of blocks, or a significant key management problem. Perhaps the hardest problem is minimizing trust. While AWS \AWSLambda runs user code with strong isolation~\cite{agache2020}, we still wish to restrict each \AWSLambda worker host to only being able to access the data it needs for the functions that have been sent to it.

The authors of Farsite~\cite{Douceur2002b, Adya2002} developed \emph{convergent encryption} as a solution to this problem. A cryptographic hash of each block (in the case of Farsite a file block, in our case a chunk of a flattened container image) is used to deterministically derive a cryptographic key that is used for encryptng the block. We follow this same scheme, but mix additional metadata into the key derivation (as described in Section \ref{sec:blast}).

The flattening process described in Section \ref{sec:blocklevel} takes each chunk, derives a key from it by computing its SHA256 digest, and then encrypts the block using AES-CTR (with the derived key). Here, AES-CTR is used with a deterministic (all zero) IV, ensuring that the same ciphertext always leads to the same plaintext. Using a deterministic IV in this context is safe, because due to the collision resistance of SHA256, a \emph{key, IV} pair is only used on for one plaintext block~\cite{dworkin2007}. A manifest of chunks is then created, containing the offset, unique key, and SHA256 hash of each chunk\footnote{It may appear attractive to use an AEAD mode like AES-GCM rather than the more expensive SHA256 in this application, but these modes do not commonly provide collision resistance against attackers who know the data key~\cite{Dodis2018}, an important property in our security scheme.}. The manifest is then encrypted, using AES-GCM, using a unique per-customer key managed by AWS Key Management Service (AWS KMS). Chunks are then named based on a function of the hash of their ciphertext, and uploaded to the backing store (AWS S3) using that name if no chunk of that name already exists. 

In our scheme, we do not encrypt the entire manifest with the customer's unique key. Instead, only the key table (the keys of each encrypted chunk) is encrypted, and the whole document is authenticated (i.e. included in the calculation of the AES-GCM tag as additional data). This allows the garbage collection process to access the list of chunks in the manifest, while having no access to the chunk keys. The size of manifests, stored in an efficient binary format, is negligible: less than 3MiB for a 16GiB container image, or 0.02\% overhead.

This approach provides a number of desirable properties:

\begin{itemize}
\item Data can be deduplicated with no sharing of keys: the keys to decrypt the customer's manifest are unique to that customer, and access to them (via AWS KMS) is only provided to the workers that that particular customer's functions are placed on. 
\item Data can be deduplicated with no coordination or special access provided to the flattening process. Flattening processes operate independently, and the only special operation they need is "upload this file to storage if it doesn't already exist". 
\item The scheme provides strong end-to-end integrity protection for chunks. Workers check the chunks they download against the MAC in the manifest, ensuring that modified ciphertexts can be detected and rejected.
\end{itemize}

\subsection{Compression}
Our system does not compress chunk plaintexts prior to encryption. This is for two reasons. First, given the network bandwidth available to our caches and workers the additional latency of decompression, and difficulty of allowing random access to compressed data, makes the latency benefit of compression marginal. Second, compression before encryption allows potential attackers to infer plaintext contents from compressed sizes, a \emph{compression side channel}. This risk, and the relatively small expected benefit, means that we decided not to implement compression (beyond trivial elision of all-zero chunks).

\subsection{Limiting Blast Radius}~\label{sec:blast}
While deduplication has value in cost and cache performance, it also adds some risks. Some popular chunks are widely referenced, meaning that anything that causes access to those chunks to break or become slow, also has a very wide impact on the system. Risks include partial (gray) failures of cache nodes, operational issues that cause unavailability of data, bugs in garbage collection, or corruption of data in the cache hierarchy. Highly popular chunks also cause hot-spotting in distributed storage. While our cryptographic scheme detects corruption and will prevent readers from seeing corrupt data, it does not correct it, and so corrupted data will become unavailable.

To solve this problem, we include a varying \emph{salt} in the key derivation step of our convergent encryption scheme. This salt value can vary in time, with chunk popularity, and with infrastructure placement (such as using different salts in different availability zones or datacenters). Otherwise-identical chunks with different salt values will end up with different keys, and therefore difference ciphertexts, and will not deduplicate against each other. By controlling the frequency with which the salt is rotated, we can continuously trade off deduplication efficiency with blast radius. Salt allows us to encapsulate the control of deduplication entirely within the chunk creation layer, without any other component needing to be aware of its decisions. Salt rotation is an operational concern, and is not needed for the security of the deduplication scheme. 

\subsection{Garbage Collection}
A key challenge of any distributed deduplication scheme is garbage collection: removing data from the backing store when it is no longer actively referenced. Garbage collecting the wrong chunk could cause wide impact across multiple customers. Our deduplication scheme does not maintain a central directory of chunk references or manifests, making exact reference counting infeasible. Past experience with distributed garbage collection has taught us that the problem is both complex (because the tree of chunk references is changing dynamically) and uniquely risky (because it is the one place in our system where we delete customer data). The approach we took to garbage collection is based on this experience.

\begin{figure}[!t]
\centering
\includegraphics[width=0.8\columnwidth]{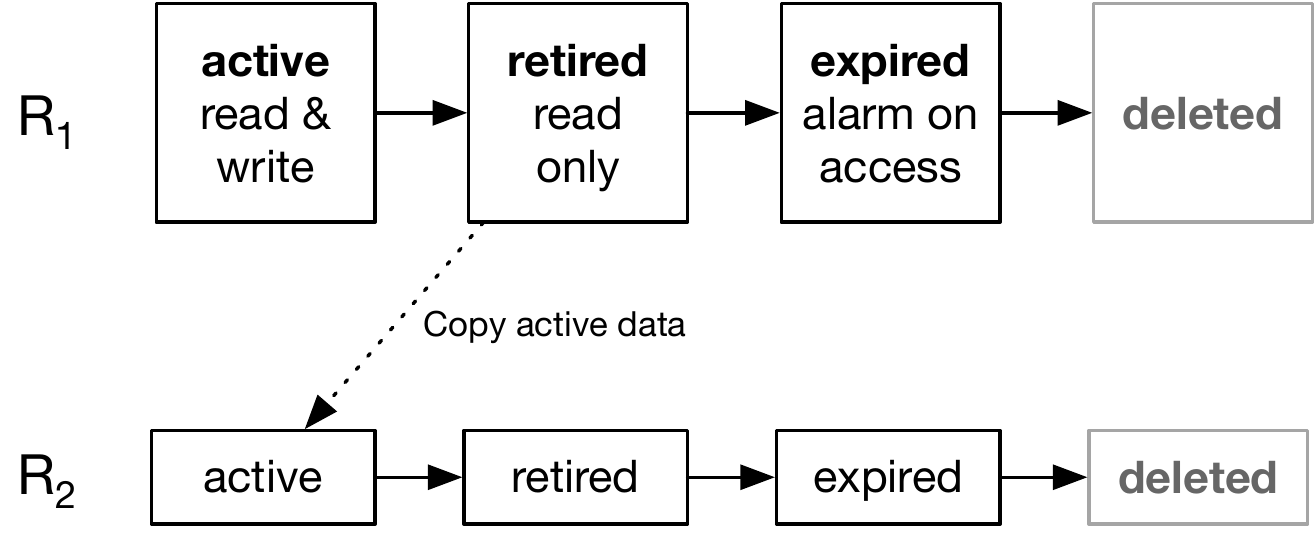}
\caption{Lifecycle of data chunks used by the generational garbage collector.}
\label{fig:gc_roots}
\end{figure}

Our approach to garbage collection is based on the concept of \emph{roots}. A \emph{root} is a self-contained manifest and chunk namespace, analogous to the \emph{root}s used in traditional garbage collection algorithms. Unlike traditional GC roots, in our system we periodically create new roots (which then get all new data), and retire old roots (after moving any still-needed data into a fresh root).%
When a customer's container image is converted, the manifest and set of chunks are placed in an \emph{active} root, for example $R_1$. An active root handles both reads and writes of data. Periodically, a new root $R_2$ is created and becomes active, while root $R_1$ enters a \emph{retired} state at which point it only serves reads of data. While $R_1$ is retired, any manifest that is still referenced in $R_1$ is migrated, along with any chunks it references, to $R_2$. Over time the manifests and chunks in $R_1$ that are in active use will be migrated to $R_2$, allowing $R_1$ to be safely deleted. This process is repeated: $R_2$ is retired and $R_3$ becomes the active root and so on. Figure \ref{fig:gc_roots} shows this lifecycle. Moving chunks along with their manifest ensures that if a manifest exists in root $R$, then all the chunks it references do to. A unique identifier for the currently active root is also included in the deduplication salt (Section \ref{sec:blast}), ensuring that newly-created chunks in the active root are not shared with previous roots.

Instead of deleting roots immediately after data migration is complete, we put them into an \emph{expired} state. In this state, data is still allowed to be read, but any attempt to access data leads to an alarm. These alarms both engage an operator and automatically stop further deletion of data. This approach allows us to robustly detect garbage collection issues (especially incomplete copying) in production, and quickly and automatically stop any data from being deleted. While this mechanism is inexact (data could be accessed after the period the root is \emph{expired}), it provides a valuable additional layer of protection against data loss. While software bugs are rare, and we test garbage collection changes carefully, multiple layers of protection against customer data loss are critical in any distributed storage system.

Having data in multiple roots does drive up storage costs, however that additional cost is palatable for \AWSLambda as customers often update their functions and a large majority of data is never migrated to a new root. The system is also capable of having multiple roots active simultaneously, which reduces the blast radius of bugs and provides the ability to roll out new garbage collection changes and algorithms to a subset of manifests and their chunks.

\section{Tiered Caching}\label{sec:cache}

When workers don't have chunks in their local cache, they attempt to pull them from a remote availability-zone-level (AZ-level) shared cache (as shown in Figure \ref{fig:high_level_arch}). If chunks aren't in this cache, workers download them from S3, and upload them into the cache. This AZ-level cache is a custom implementation of a fairly standard design: chunks are fetched over HTTP2, data storage is two-tiered with an in-memory tier for hot chunks and a flash tier for colder chunks, and eviction is LRU-k~\cite{ONeil1993} (a scan-resistant variant of Least Recently Used). Chunks are distributed to the AZ-level cache using a variant of a consistent hashing~\cite{Karger1997} scheme, with optimizations to improved load spreading (similar to the approach of Chen et al~\cite{Chen2019}). The caching tier improves fetch performance considerably: from the worker's perspective, a hit on the AZ-level cache takes a median time of $550\mu s$, versus $36ms$ for a fetch from the origin in S3 (99.9th percentile $3.7ms$ versus $175ms$).

\begin{figure}[!t]
\centering
\includegraphics[width=0.8\columnwidth]{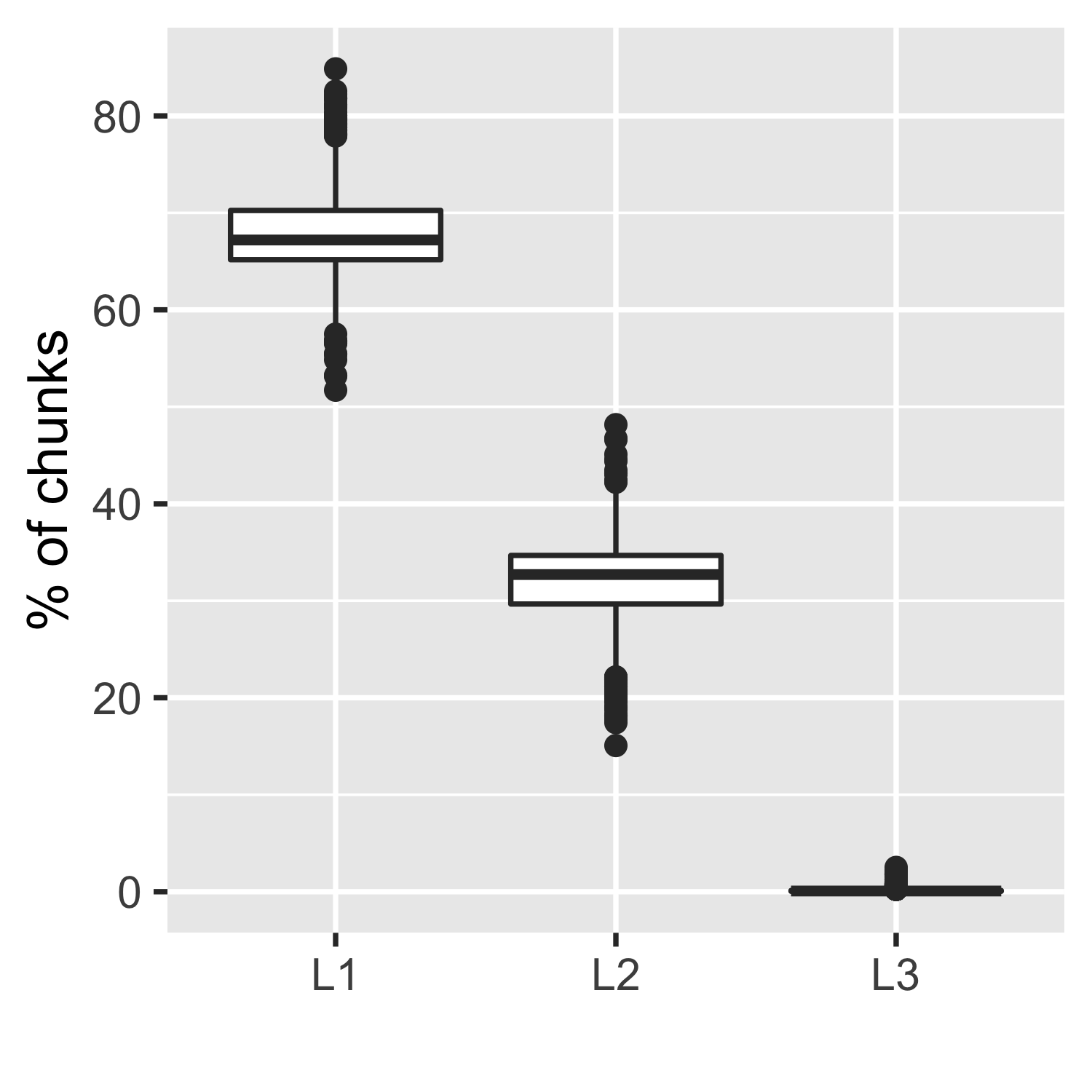}
\caption{One week of hit rates on each of the cache tiers: on-worker (L1), distributed in-AZ (L2), and backing store (L3)}
\label{fig:chunk_locations}
\end{figure}

Figure \ref{fig:chunk_locations} shows the effectiveness of these three cache tiers. Over a week of production usage in one large AWS region, a median of 67\% of chunks were loaded from the on-worker cache, 32\% from the AZ-level distributed cache, and the remaining 0.06\% from the backing store. 

\begin{figure}[!t]
\centering
\includegraphics[width=0.8\columnwidth]{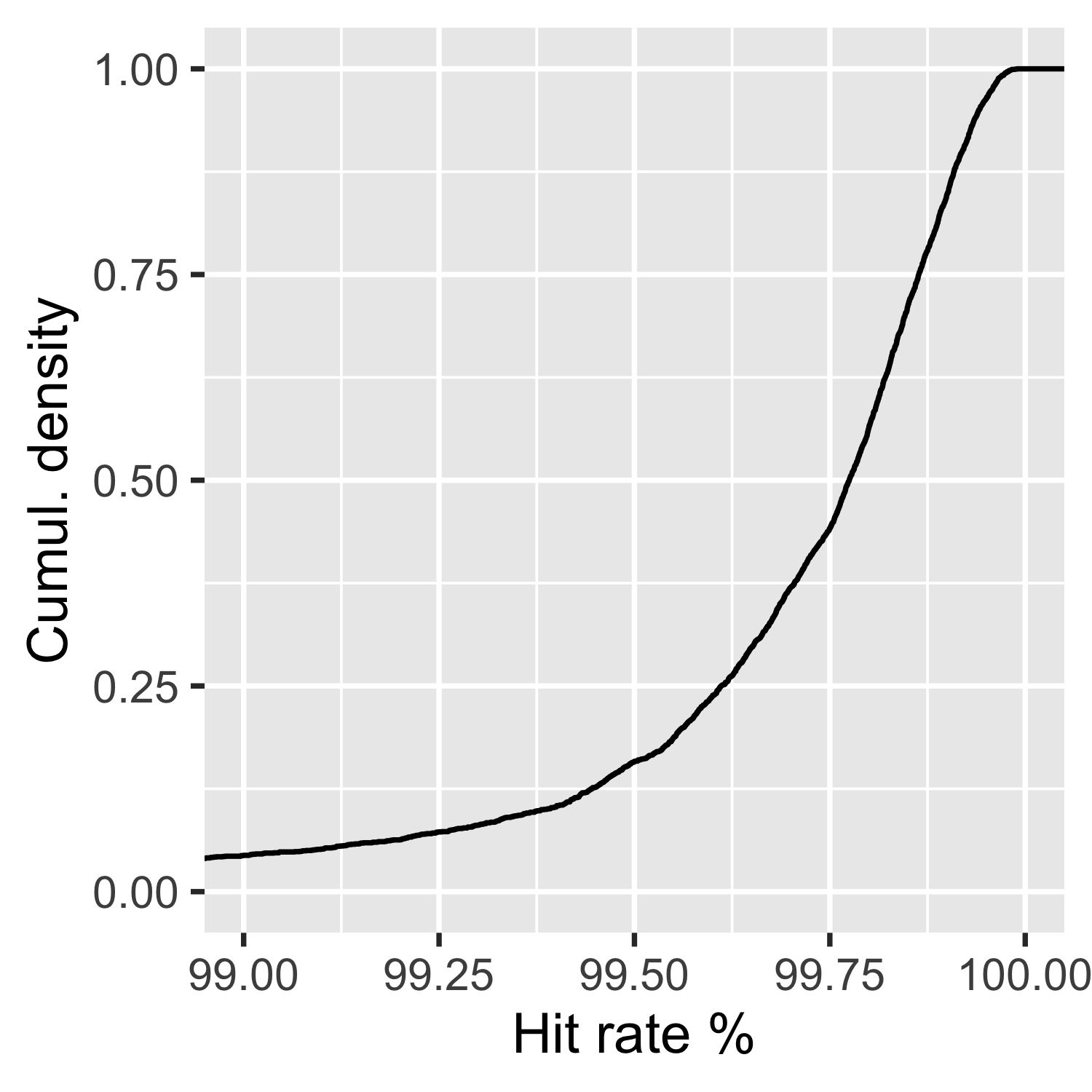}
\caption{Empirical CDF of in-AZ cache hit rate}
\label{fig:hit_rate}
\end{figure}

The per-worker cache has a median hit rate of 67\%, and a 10th percentile low hit rate over the week in question of 65\%. The in-AZ cache is even more effective, with a median hit rate of 99.9\% and 10th percentile low hit rate over the week of 99.4\%. Figure \ref{fig:hit_rate} shows the empirical CDF of the hit rate of the in-AZ cache over the week, measured in one-minute buckets across one at-scale production availability zone. The left tail of the distribution is associated with large spikes in traffic to newly created functions. We are evaluating priming the in-AZ caches during the chunk creation process to flatten this left tail and further improve hit rates, primarily with the goal of reducing load-time latency for new functions.

\subsection{Optimizing for Tail Latency}

While data in the AZ-level cache is not required to be durable (durability is ensured using \SThree as the origin), a simple unreplicated cache scheme (where each object is stored in a single node) didn't meet our needs for three reasons.

\begin{description}
	\item[Tail latency] A single slow cache server can cause widespread impact. Slowness could be caused by congestion at the host or in the network, or by partial hardware or software failure.
	\item[Hit Rate Drops] Having each item cached in a single server means that the hit rate drops if that server fails, or is taken down for deployment.
	\item[Throughput Bounds] Having each item cached in a single server means that the bandwidth available to fetch the object is bounded by a single server's bandwidth.
\end{description}

Of these, tail latency is the largest practical concern. Our experience operating these types of systems suggests that debugging slowness and partial failure is much harder than debugging outright failure. Even if this slowness is only in the long tail, it still matters in production because each container start needs to fetch a large number of chunks. For example, a start which fetches 1000 chunks will experience the 99.9th percentile tail latency of the cache on 63\% of tasks. The difference is material: in one deployment of the cache we observe a median client-measured latency of $500 \mu s$, and a 99.9th percentile latency of $4ms$.

\begin{figure}[!t]
\centering
\includegraphics[width=\columnwidth]{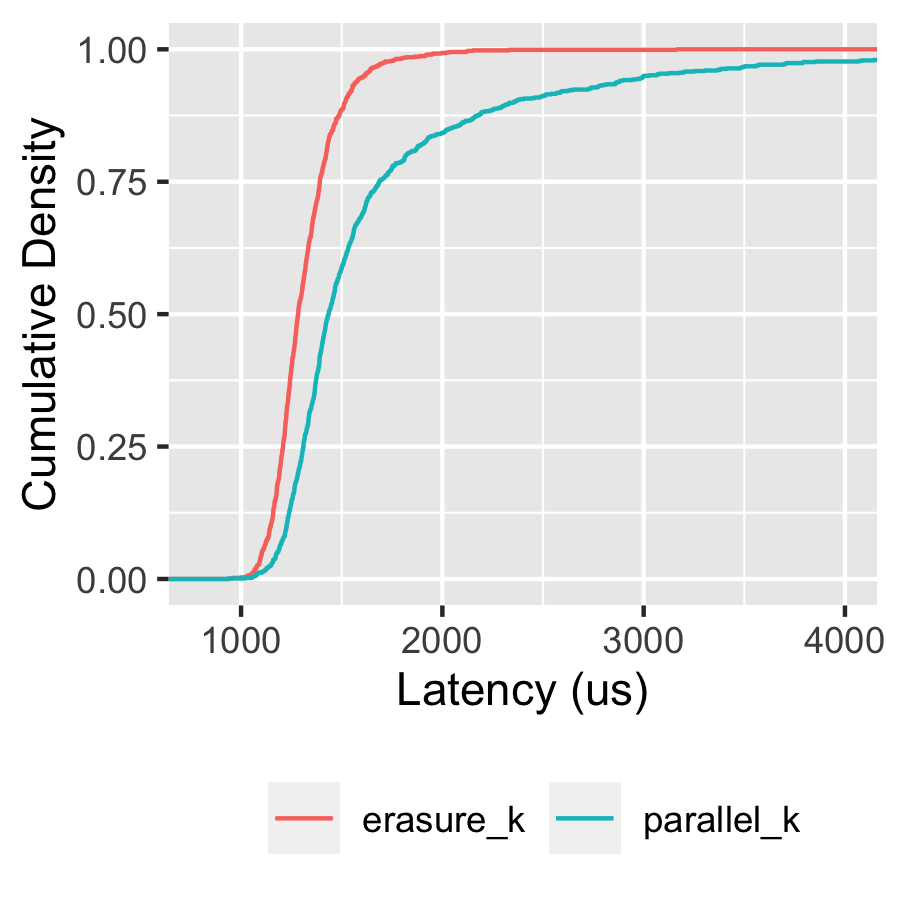}
\caption{Comparative empirical CDFs of client-side latency of 4-of-4 parallel cache load, versus 4-of-5 erasure coded cache load.}
\label{fig:erasure_cdf}
\end{figure}

Replication, combined with redundant requests is a well-established~\cite{Wu2015, Gardner2015, Ashish2013} technique to drive down tail latency, and would also solve our throughput and hit-rate problems. Unfortunately, replication increases costs proportionally to the replication factor, an important concern in a primarily in-memory cache. Instead, we chose erasure coding, following a similar scheme to EC-Cache~\cite{Rashmi2016}. Erasure coding is not widely used in caches, but provides compelling solutions for all three of our concerns. When a worker misses the cache, it fetches the chunk it needs from the origin, then uploads erasure-coded stripes of that chunk into the cache. When a worker needs to fetch a chunk, it requests more stripes than are strictly needed to reconstruct the chunk, and then reconstructs the chunk as soon as enough stripes are returned. Our current production deployment uses a \emph{4 of 5} code, achieving 25\% storage overhead, and a 25\% increase in request rate in exchange for a significant decrease in tail latency. Figure \ref{fig:erasure_cdf} compares the empirical latency CDF of the \emph{4 of 5} code versus a hypothetical \emph{4 of 4} scheme using latency measurements from one deployment of our production system.

This scheme prevents any drop in hit rate from occurring when cache nodes fail, or are taken down for deployment. A common approach in similar systems is to use retries to hide the effects of deployments and failed nodes, an approach which is known to lead to metastable failure modes in large systems~\cite{Bronson2021, Huang2022}. Erasure coding allows us to achieve a similar level of resiliency while performing the same amount of work in success and failure cases (a design philosophy we call \emph{constant work}~\cite{Colm2020}).

\subsection{Stability and Metastability}\label{sec:metastability}
Caches with high hit rates, such as ours, are desirable from a latency and efficiency perspective, but have a hidden downside. If the cache becomes empty (such as due to power loss or operational issue), or the hit rate suddenly drops (such as due to a change in customer behavior), the downstream services can see significantly more traffic than they are used to. In the case of our cache, with an end-to-end hit rate typically exceeding 99.8\%, this downstream traffic increase could be up to 500 times normal. \SThree is an extremely scalable backing store, and can tolerate the full uncached load. However, the increased latency leads to higher concurrency demand from customer's applications (due to Little's Law~\cite{little1961}), and therefore higher demand for new \AWSLambda slots, increasing load and changing the size and composition of the system's working set. This can lead to metastable behavior~\cite{Bronson2021, Huang2022, Brooker2019}, where the system isn't able to refill the cache when it is empty\footnote{Related effects have been observed in computer systems since at least the 1960s. In the 1968 paper `The Working Set Model for Program Behavior'~\cite{Denning1968}, Peter J Denning observed a similar effect in paging systems: \begin{quote} This can create a self-intensifying crisis. Programs, deprived of still-needed pages, generate a plethora of page faults; the resulting traffic of returning pages displaces still other useful pages, leading to more page faults, and so on.\end{quote}}.

We have built mitigations for this risk into higher layers of \AWSLambda. Primarily, the system is designed to be concurrency-limited. When container starts slow down and the number of concurrent tasks exceeds this limit, new starts are rejected until in-flight ones complete. We also actively test the system's ability to \emph{cold start} from an empty cache at the maximum concurrency. This testing allows us to be confident that the system is able to restart from a cold cache, or tolerate workload changes that significantly reduce hit rate.

\subsection{Cache Eviction and Sizing}
Traditional cache replacement policies like Least Recently Used (LRU) and First In First Out (FIFO) are simple and easy to implement, but have a significant downside for this application: a lack of \emph{scan resistance}. In our case, this means that a large number of infrequently used functions starting up can replace all the hot entries in the cache with recently-used entries belonging to those functions, dropping cache hit rates for more frequently-used entries, and filling the cache with entries that will never be read again. This happens periodically in our environment, driven by weekly, daily, and hourly spikes of periodic \emph{cron job} functions. These functions are large in number, but each runs at a low scale (typically only using one sandbox), making caching their chunks relatively unimportant. To avoid the hit-rate drops caused by this periodic work, we use the LRU-k~\cite{ONeil1993} eviction algorithm, which tracks the last $k$ times an item in the cache was used, rather than only the most recent time.

Eviction and hit rates are also related to the size of our local and AZ-level caches. Following the logic of Gray and Putzolu's classic \emph{Five Minute Rule}~\cite{gray1987}, the minimum desirable cache size is the one that makes the cost of cache retention equal to the cost of fetching chunks from S3. However, because our cache is not only aimed at reducing costs but also improving customer-observed latency, we also set a hit rate goal and increase the cache size if we fall below that goal. The total cache size, then, is the larger of the size needed to achieve our hit rate goal, and the size needed to optimize costs.

\section{Implementation and Production Experience}

We built the local agent (the FUSE implementation that backs the sparse block device for each MicroVM), the worker-local cache, and the remote cache server in the Rust programming language. We used the \emph{tokio} runtime, and \emph{reqwest} and \emph{hyper} for HTTP. At the time we started this project, the invoke path of \AWS \AWSLambda includes components written in Java, Go, C, and Rust. We chose Rust because of our good experiences with the Rust components we had built in the past, especially around performance and stability, and have again been happy with our choice of Rust, encountering no major production bugs in the libraries we chose. We were also attracted to Rust because of the successes other \AWS teams (such as the \Amazon \SThree team~\cite{Bornholt2021}) have had applying formal methods to verify code correctness in Rust, even with non-expert programmers.

One interesting stumbling block with Rust (version 1.46.0, current at the time of implementation) is brittle optimization, especially autovectorization, of hotspots. Unsurprisingly, we found that the parity calculations we use for erasure coding are nearly 5x faster when performed 64 bytes at a time (with AVX512) or 32 bytes at a time (with AVX or NEON) than when performed 8 bytes at a time, and 10x faster than when performed byte-at-a-time. Unfortunately, the naive Rust loop emitted the byte-at-a-time code (as shown in Listing \ref{list:parity_x86}), despite the compiler being capable of autovectorization. Small changes to the code would change autovectorization behavior, even changes outside the function of interest. Reluctant to move to assembly for this code, we finally settled on the code in Listing \ref{list:autovec}, which robustly emits appropriately unrolled AVX, AVX512, or vectorized ARM code depending on the target platform. Seemingly small changes to this function (such as removing the \emph{assert}, changing any of the assignments, or allowing it be inlined) cause autovectorization to be disabled. This is a small issue with Rust, and one that we expect to be improved in future compiler versions.

\begin{lstfloat}
\begin{verbatim}
  0.08 |350:   cmp    %rax,%rsi                                                                                                                                        
       |     ↓ jae    3f4                                                                                                                                              
 49.18 |       movzbl (%rdi,%rsi,1),%ebx                                                                                                                               
  0.13 |       xor    %bl,(%rcx,%rsi,1)                                                                                                                                
 50.52 |       lea    0x1(%rsi),%rbp                                                                                                                                   
  0.08 |       mov    %rbp,%rsi                                                                                                                                        
       |       cmp    %rax,%rbp                                                                                                                                        
       |     ↑ jb     350
\end{verbatim}
\caption{Naive byte-by-byte x86 assembly code as emitted by the Rust compiler for straightforward loop implementation (with annotations by \emph{perf} showing percent of runtime). Note significant missed opportunities for optimizations like vectorization and loop unrolling.}
\label{list:parity_x86}
\end{lstfloat}

\begin{lstfloat}
\begin{verbatim}
#[inline(never)]
fn parity(target: &mut [u8], source: &[u8]) {
    assert_eq!(source.len(), target.len());
    let len = target.len();
    let _ = target[len-1];
    let _ = source[len-1];

    for i in 0..len {
        target[i] ^= source[i];
    }
}
\end{verbatim}
\caption{Implementation of parity calculation in Rust, showing extra lines needed for reliable autovectorization.}
\label{list:autovec}
\end{lstfloat}

On the other hand, the Rust ecosystem's support for build-time microbenchmarks (such as with the \emph{criterion} crate) makes it fast and easy to iterate on this type of performance work, and even assert at build time that autovectorization has succeeded (effectively stopping regressions from entering production). This is a significant boon in a cloud environment, where performance regressions can cause production outages, and performance is tied to both cost and carbon efficiency.

\subsection{Latency and Multimodality}

As with any storage system, performance was an important goal for the design and implementation of our snapshot chunk loading system. While throughput, CPU efficiency, and other bandwidth measures contribute to the cost of running the system, its scale-out nature make latency and scalability the most important factors of performance. The local agent and on-worker caches trivially scale out, due to the fact that they do not communicate off their worker, except in interacting with S3 (to pull chunks from the origin), and the L2 AZ-level cache.

\begin{figure}[!t]
\centering
\includegraphics[width=0.8\columnwidth]{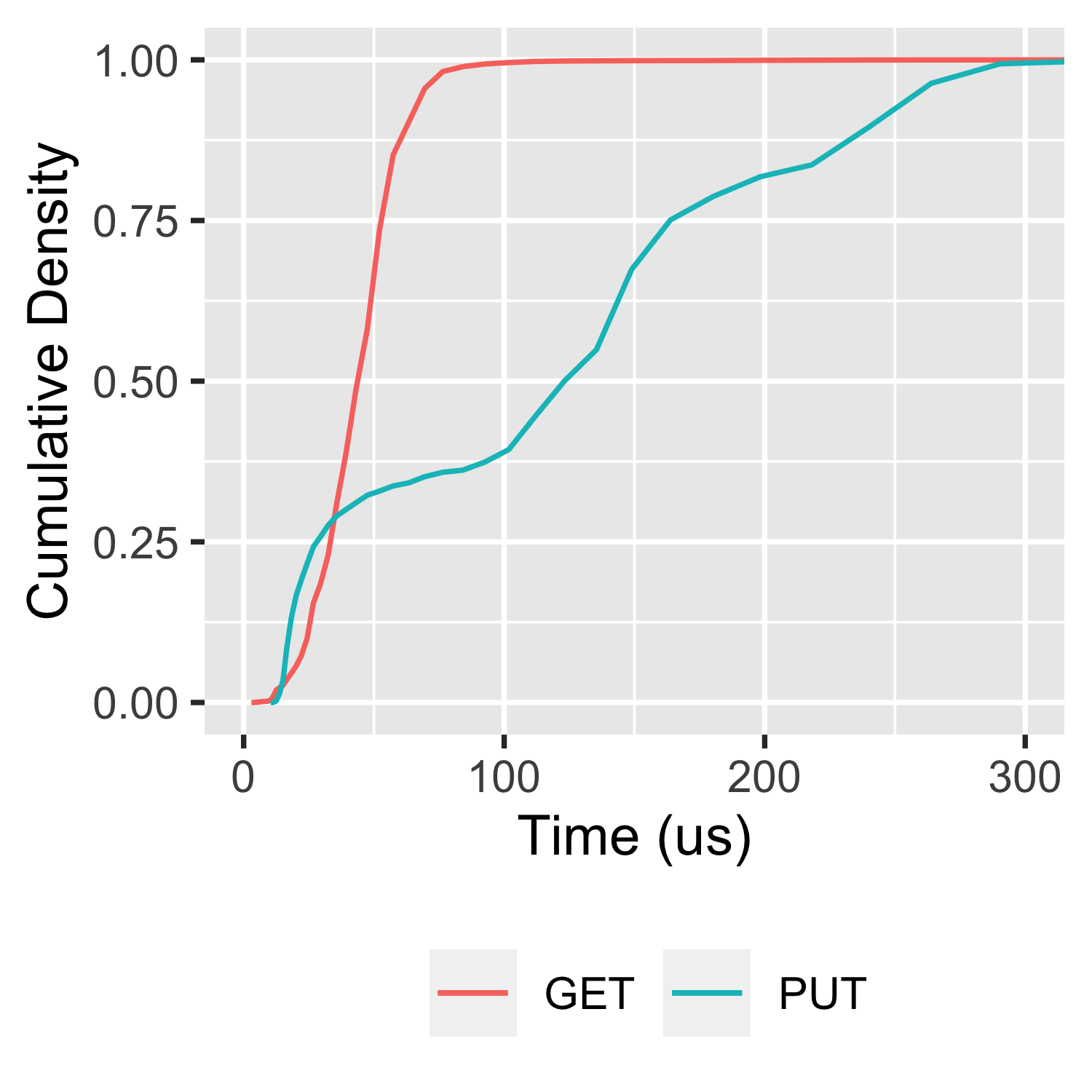}
\caption{Empirical CDF of server-side measured latency of the L2 cache server}
\label{fig:l2_server_latency}
\end{figure}

Figure \ref{fig:l2_server_latency} shows the latency for GETs and PUTs on this cache, measured from the server side, across all of the cache nodes in a production deployment over the course of one week. Each GET or PUT is of a 512kB chunk. As discussed in Section \ref{sec:cache}, the L2 cache is a flash-based cache with a significant local memory tier (about 10\% of cache size). GET latency is very consistent, with a median of below $50 \mu s$. PUT latency is less consistent, with some multi-modality apparently caused by writeback behavior on the cache host. Despite this multi-modality, performance is still excellent, with a median latency of $125 \mu s$, a 99th percentile latency below $300 \mu s$, and a 99.99th percentile of $413 \mu s$\footnote{Having a 99.99th percentile at less than 4x the median is a very desirable property, and difficult to achieve with garbage collected languages like Java and Go}. When building this cache server, we chose HTTP2 as a wire protocol for convenience with the intention of replacing it with an efficient binary protocol later. In production, we've found the overhead of HTTP (implemented with \emph{hyper} and \emph{reqwest}) so low that we have not yet been motivated to replace the protocol.

\begin{figure}[!t]
\centering
\includegraphics[width=0.8\columnwidth]{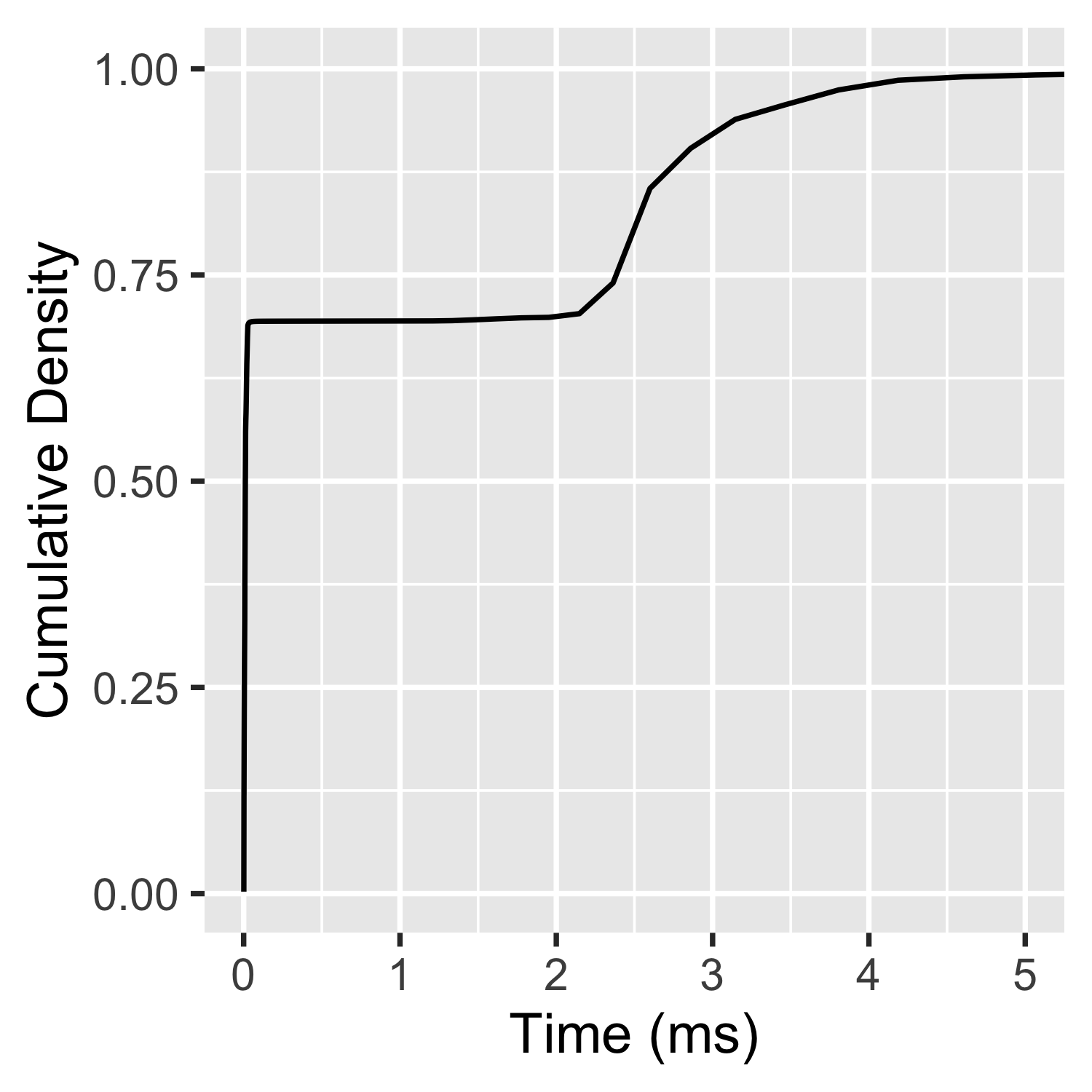}
\caption{Empirical CDF of end-to-end read latency observed at the local agent (FUSE implementation).}
\label{fig:e2e_latency}
\end{figure}

Figure \ref{fig:e2e_latency} shows the end-to-end latency for returning a read from the perspective of the local agent (that is the FUSE implementation). This doesn't show the end-to-end IO latency experienced by guests, because it's from the perspective of the worker and does not include the (significant) hit rate on the page cache maintained by the MicroVM guest's kernel, and read-ahead performed by the guest to populate that cache. Like the L2 server latency, this end-to-end latency shows significant multi-modality: a mode below $100 \mu s$ which represent local cache hits, a mode around $2.75ms$ which represent L2 hits (and the subsequent work like decryption), and mode (trimmed from the graph) showing rare fetches from the origin (see Figure \ref{fig:chunk_locations} for the relative frequencies of these modes). We are working on an optimized cryptographic scheme which reduces the latency of decryption.

Multi-modality like this is the norm in storage systems, but presents a few practical challenges to operators. First, as discussed in Section \ref{fig:chunk_locations} a small change in the relative frequencies of each mode can significantly change the mean latency observed by clients (and so change the concurrency and throughput of the system). Second, latency percentiles and trimmed means are the summary statistics most commonly used by operators at \AWS, and they tend to obscure multi-modality. Plots like empirical CDFs (eCDFs, as presented here) can be valuable, but don't show change-over-time as time series of summary statistics do. We have experimented with heat maps, day-over-day eCDFs, and others, but have yet to find a succinct way to present these data to operators. Third, multi-modality makes the decision of where to spend optimization resources more complex. Which mode should the team work to improve? Or should they work to reduce the relative frequencies of higher modes?

\subsection{Production experience with FUSE}
Our experiences with FUSE match those reported by Vangoor et al~\cite{Vangoor2017}, showing relatively little throughput overhead when well tuned. However, we have found that the choice to use FUSE to present a file which is then subsequently used as a block device by Firecracker's virtio-blk implementation, has introduced significant overhead. When an application running in a MicroVM reads a new chunk, control is passed to the guest kernel, then Firecracker, then the host kernel's FUSE layer, then the local agent, before flowing back through the same path. This introduces context switch overhead, but more importantly requires four different threads to be scheduled by the host kernel's scheduler. This introduces inefficiency in steady state, and significant jitter under load. We are moving away from FUSE for this application, primarily due to this effect. Our new implementation uses \emph{userfaultfd} and \emph{mmap}, removing two layers from the architecture.

We don't regret starting with FUSE. It provided a convenient interface, a clear security and operational isolation story, and allowed a team without deep systems-level programming experience to build an acceptably high performance system.

\section{Related work}\label{sec:related}

Mirroring the rise in popularity of serverless and containers accelerated container loading has been a highly active area of research, and industry implementation, over the last decade. Before that, accelerating VM loading through faster disc image movement was an active area of research. For example, Frisbee~\cite{Hibler2003} in 2003. Amazon EC2 has taken advantage of common data to accelerate VM image loading, through tracking lineage of EBS snapshot chunks~\cite{Olson2021}, since 2009. With Slacker~\cite{Harter2016} Harter et al studied access patterns in container loading, and presented a system which takes advantage of these patterns by performing layer-level lazy loading. Starlight \cite{Chen2022} takes a fairly similar filesystem-orientated approach, optimized for loading at the edge where minimizing round-trips to the datacenter is a significant contributor to performance. eStargz \cite{Tokunaga2020} extends common container image formats to make lazy loading at the layer level more efficient, building on the approach of Google's CRFS.

DADI \cite{Li2020} uses a block-level approach fairly similar to our own, but with a peer-to-peer approach rather than a dedicated cache layer, and without the ability to deduplicate as widely as our system is able to. FaaSNet \cite{Wang2021} approaches a similar problem to the one we were solving, but works on the layer level (rather than flattening images as we do), and does not appear to perform deduplication. Cntr \cite{Thalheim2018} and Yolo \cite{nguyen2019} take the approach of breaking down container images into different classes of data, some needed urgently on start up and some likely to be accessed less urgently. This explicit approach may be more efficient than the simple block-based approach, but also requires a deeper introspection of the contents of the container. Wharf \cite{Chao2018} and CFS \cite{Liu2019} take the distributed filesystem approach, showing that can significantly improve loading performance at the cost of increased coordination between containers.

Accelerating storage performance and loading with deduplication has an even longer history, for example in 2001 with Muthitacharoen et al~\cite{Muthitacharoen2001} and 2002 with Venti~\cite{Quinlan2002}, and Farsite~\cite{Adya2002}.
\section{Conclusion}

We present \AWS \AWSLambda's solution for accelerated loading of container images, and approach that combines deduplication, erasure coding, tiered caching, userspace filesystems, and convergent encryption. We have operated this system for several years, and are extending its use into other areas of \AWS. While our solution on the surface appears to have a lot of moving parts, it is optimized for what we believe to be the realities of building massive scale cloud systems: failures are frequent, failures are often partial and complex, and security is the top priority.

\subsection{Broader Lessons and Future Work}
While \AWSLambda's snapshot loading infrastructure is a specialized system for a rather specialized application, we believe that there are some broader lessons from our experiences that apply to the systems community as a whole.

\begin{itemize}
\item Containers are most popularly used by \AWSLambda customers as ``static linking in the large'' dependency closures. Customers want to build, test, and deploy a function with all its dependencies in one atomic unit, but traditional static linking is either unavailable or inconvenient. However, containers are also highly inefficient in this context, necessitating the deduplication and sparse loading we describe here. We believe that there is a significant need for a lighter-weight dependency closure mechanism, which comes closer to traditional static linking in the size of the artifacts that it creates.

\item Caches reduce costs, improve latency, and reduce load on durable storage, and are a critical component of nearly any stateful system. However, they also introduce risks such as metastable failures (due to unexpectedly empty caches, or sudden shifts in workloads), and challenges for users like multi-modal latency distributions. While work such Yang et al~\cite{Yang2020}, and Huang et al~\cite{Huang2022} have made steps towards deeply understanding these effects, we believe that significantly more work is needed to understand the dynamic behaviors of caching in large systems, and to develop patterns to mitigate the risks of caches.

\item MicroVMs provide an isolation mechanism which is nearly as lightweight as containers, or even processes~\cite{agache2020, Manco2017}, while providing additional interfaces for plugging in both local and distributed operating system logic. MicroVMs provide a powerful new tool in the operating system researcher's or builder's toolbox. We believe that operating system support for virtualization, and virtualization support for applications, operating systems, and databases are ripe areas of research which are not yet receiving sufficient attention.
\end{itemize}

Our future work is focused on optimizing the system further for cost, performance, and especially customer-experienced cold-start latency. This same system is used in Lambda SnapStart, a feature of AWS Lambda which reduces cold-start latency using memory snapshots, to store and load memory snapshot contents. That use-case is especially latency sensitive, motivating significant investments in both average case and tail latency. We expect this work to include optimizing cache retention and data placement policies, optimizing client and server performance, and completing the migration from FUSE to \emph{userfaultfd}.

\section*{Acknowledgements}

Any system of this size requires a team to build and operate, and in this case we're deeply thankful to the AWS Lambda team for their work and contributions. Holly Mesrobian, David R. Richardson, Ajay Nair, and David Nasi were instrumental in supporting this work. Shay Gueron, Osman Surkatty, and Derek Manwaring helped ground our cryptographic ambitions, and provided valuable feedback.

\bibliographystyle{plainurl}
\bibliography{paper}

\begin{thebibliography}{10}

\bibitem{OCISpec}
Oci image format specification.
\newblock Accessed: 2022-04-15.
\newblock URL: \url{https://github.com/opencontainers/image-spec}.

\bibitem{Adya2002}
Atul Adya, William~J. Bolosky, Miguel Castro, Gerald Cermak, Ronnie Chaiken,
  John~R. Douceur, Jon Howell, Jacob~R. Lorch, Marvin Theimer, and Roger~P.
  Wattenhofer.
\newblock {FARSITE}: Federated, available, and reliable storage for an
  incompletely trusted environment.
\newblock In {\em 5th Symposium on Operating Systems Design and Implementation
  (OSDI 02)}, Boston, MA, December 2002. USENIX Association.
\newblock URL:
  \url{https://www.usenix.org/conference/osdi-02/farsite-federated-available-and-reliable-storage-incompletely-trusted-environment}.

\bibitem{agache2020}
Alexandru Agache, Marc Brooker, Alexandra Iordache, Anthony Liguori, Rolf
  Neugebauer, Phil Piwonka, and Diana-Maria Popa.
\newblock Firecracker: Lightweight virtualization for serverless applications.
\newblock In {\em 17th $\{$USENIX$\}$ Symposium on Networked Systems Design and
  Implementation ($\{$NSDI$\}$ 20)}, pages 419--434, February 2020.

\bibitem{Bornholt2021}
James Bornholt, Rajeev Joshi, Vytautas Astrauskas, Brendan Cully, Bernhard
  Kragl, Seth Markle, Kyle Sauri, Drew Schleit, Grant Slatton, Serdar Tasiran,
  Jacob Van~Geffen, and Andrew Warfield.
\newblock Using lightweight formal methods to validate a key-value storage node
  in amazon s3.
\newblock In {\em Proceedings of the ACM SIGOPS 28th Symposium on Operating
  Systems Principles}, SOSP '21, page 836–850, New York, NY, USA, 2021.
  Association for Computing Machinery.
\newblock \href {https://doi.org/10.1145/3477132.3483540}
  {\path{doi:10.1145/3477132.3483540}}.

\bibitem{Bronson2021}
Nathan Bronson, Abutalib Aghayev, Aleksey Charapko, and Timothy Zhu.
\newblock Metastable failures in distributed systems.
\newblock In {\em Proceedings of the Workshop on Hot Topics in Operating
  Systems}, HotOS '21, page 221–227, New York, NY, USA, 2021. Association for
  Computing Machinery.
\newblock \href {https://doi.org/10.1145/3458336.3465286}
  {\path{doi:10.1145/3458336.3465286}}.

\bibitem{Brooker2019}
Marc Brooker.
\newblock Some risks of coordinating only sometimes.
\newblock In {\em High Performance Transaction Systems 2019 (HPTS'19)},
  November 2019.

\bibitem{Chen2019}
John Chen, Ben Coleman, and Anshumali Shrivastava.
\newblock Revisiting consistent hashing with bounded loads, 2019.
\newblock URL: \url{https://arxiv.org/abs/1908.08762}, \href
  {https://doi.org/10.48550/ARXIV.1908.08762}
  {\path{doi:10.48550/ARXIV.1908.08762}}.

\bibitem{Chen2022}
Jun~Lin Chen, Daniyal Liaqat, Moshe Gabel, and Eyal de~Lara.
\newblock Starlight: Fast container provisioning on the edge and over the
  {WAN}.
\newblock In {\em 19th USENIX Symposium on Networked Systems Design and
  Implementation (NSDI 22)}, pages 35--50, Renton, WA, April 2022. USENIX
  Association.
\newblock URL:
  \url{https://www.usenix.org/conference/nsdi22/presentation/chen-jun-lin}.

\bibitem{Denning1968}
Peter~J. Denning.
\newblock The working set model for program behavior.
\newblock {\em Commun. ACM}, 11(5):323–333, may 1968.
\newblock \href {https://doi.org/10.1145/363095.363141}
  {\path{doi:10.1145/363095.363141}}.

\bibitem{Dodis2018}
Yevgeniy Dodis, Paul Grubbs, Thomas Ristenpart, and Joanne Woodage.
\newblock Fast message franking: From invisible salamanders to encryptment.
\newblock In {\em Advances in Cryptology – CRYPTO 2018: 38th Annual
  International Cryptology Conference, Santa Barbara, CA, USA, August 19–23,
  2018, Proceedings, Part I}, page 155–186, Berlin, Heidelberg, 2018.
  Springer-Verlag.
\newblock \href {https://doi.org/10.1007/978-3-319-96884-1_6}
  {\path{doi:10.1007/978-3-319-96884-1_6}}.

\bibitem{Douceur2002b}
John~R. Douceur, Atul Adya, William~J. Bolosky, Dan Simon, and Marvin Theimer.
\newblock Reclaiming space from duplicate files in a serverless distributed
  file system.
\newblock In {\em Proceedings of the 22 Nd International Conference on
  Distributed Computing Systems (ICDCS'02)}, ICDCS '02, page 617, USA, 2002.
  IEEE Computer Society.

\bibitem{dworkin2007}
Morris~J Dworkin.
\newblock {\em NIST SP 800-38D. recommendation for block cipher modes of
  operation: Galois/counter mode (gcm) and gmac}.
\newblock National Institute of Standards \& Technology, 2007.

\bibitem{Gardner2015}
Kristen Gardner, Samuel Zbarsky, Sherwin Doroudi, Mor Harchol-Balter, and Esa
  Hyytia.
\newblock Reducing latency via redundant requests: Exact analysis.
\newblock {\em SIGMETRICS Perform. Eval. Rev.}, 43(1):347–360, jun 2015.
\newblock \href {https://doi.org/10.1145/2796314.2745873}
  {\path{doi:10.1145/2796314.2745873}}.

\bibitem{gray1987}
Jim Gray and Franco Putzolu.
\newblock The 5 minute rule for trading memory for disc accesses and the 10
  byte rule for trading memory for cpu time.
\newblock In {\em Proceedings of the 1987 ACM SIGMOD international conference
  on Management of data}, pages 395--398, 1987.

\bibitem{Harter2016}
Tyler Harter, Brandon Salmon, Rose Liu, Andrea~C. Arpaci-Dusseau, and Remzi~H.
  Arpaci-Dusseau.
\newblock Slacker: Fast distribution with lazy docker containers.
\newblock In {\em 14th USENIX Conference on File and Storage Technologies (FAST
  16)}, pages 181--195, Santa Clara, CA, February 2016. USENIX Association.
\newblock URL:
  \url{https://www.usenix.org/conference/fast16/technical-sessions/presentation/harter}.

\bibitem{Hibler2003}
Mike Hibler, Leigh Stoller, Jay Lepreau, Robert Ricci, and Chad Barb.
\newblock Fast, scalable disk imaging with frisbee.
\newblock In {\em 2003 USENIX Annual Technical Conference (USENIX ATC 03)}, San
  Antonio, TX, June 2003. USENIX Association.
\newblock URL:
  \url{https://www.usenix.org/conference/2003-usenix-annual-technical-conference/fast-scalable-disk-imaging-frisbee}.

\bibitem{Huang2022}
Lexiang Huang, Matthew Magnusson, Abishek~Bangalore Muralikrishna, Salman
  Estyak, Rebecca Isaacs, Abutalib Aghayev, Timothy Zhu, and Aleksey Charapko.
\newblock Metastable failures in the wild.
\newblock In {\em 16th USENIX Symposium on Operating Systems Design and
  Implementation (OSDI 22)}, pages 73--90, Carlsbad, CA, July 2022. USENIX
  Association.
\newblock URL:
  \url{https://www.usenix.org/conference/osdi22/presentation/huang-lexiang}.

\bibitem{Ionescu2022}
Vlad Ionescu.
\newblock Scaling containers on aws in 2022.
\newblock Accessed: 2022-04-15.
\newblock URL:
  \url{https://www.vladionescu.me/posts/scaling-containers-on-aws-in-2022/}.

\bibitem{Karger1997}
David Karger, Eric Lehman, Tom Leighton, Rina Panigrahy, Matthew Levine, and
  Daniel Lewin.
\newblock Consistent hashing and random trees: Distributed caching protocols
  for relieving hot spots on the world wide web.
\newblock In {\em Proceedings of the Twenty-Ninth Annual ACM Symposium on
  Theory of Computing}, STOC '97, page 654–663, New York, NY, USA, 1997.
  Association for Computing Machinery.
\newblock \href {https://doi.org/10.1145/258533.258660}
  {\path{doi:10.1145/258533.258660}}.

\bibitem{Li2020}
Huiba Li, Yifan Yuan, Rui Du, Kai Ma, Lanzheng Liu, and Windsor Hsu.
\newblock {DADI}: {Block-Level} image service for agile and elastic application
  deployment.
\newblock In {\em 2020 USENIX Annual Technical Conference (USENIX ATC 20)},
  pages 727--740. USENIX Association, July 2020.
\newblock URL:
  \url{https://www.usenix.org/conference/atc20/presentation/li-huiba}.

\bibitem{little1961}
John~DC Little.
\newblock A proof for the queuing formula: L= $\lambda$ w.
\newblock {\em Operations research}, 9(3):383--387, 1961.

\bibitem{Liu2019}
Haifeng Liu, Wei Ding, Yuan Chen, Weilong Guo, Shuoran Liu, Tianpeng Li, Mofei
  Zhang, Jianxing Zhao, Hongyin Zhu, and Zhengyi Zhu.
\newblock Cfs: A distributed file system for large scale container platforms.
\newblock In {\em Proceedings of the 2019 International Conference on
  Management of Data}, SIGMOD '19, page 1729–1742, New York, NY, USA, 2019.
  Association for Computing Machinery.
\newblock \href {https://doi.org/10.1145/3299869.3314046}
  {\path{doi:10.1145/3299869.3314046}}.

\bibitem{Colm2020}
Colm MacCárthaigh.
\newblock Reliability, constant work, and a good cup of coffee, 2020.
\newblock URL:
  \url{https://aws.amazon.com/builders-library/reliability-and-constant-work/}.

\bibitem{Manco2017}
Filipe Manco, Costin Lupu, Florian Schmidt, Jose Mendes, Simon Kuenzer, Sumit
  Sati, Kenichi Yasukata, Costin Raiciu, and Felipe Huici.
\newblock My vm is lighter (and safer) than your container.
\newblock In {\em Proceedings of the 26th Symposium on Operating Systems
  Principles}, SOSP '17, page 218–233, New York, NY, USA, 2017. Association
  for Computing Machinery.
\newblock \href {https://doi.org/10.1145/3132747.3132763}
  {\path{doi:10.1145/3132747.3132763}}.

\bibitem{Muthitacharoen2001}
Athicha Muthitacharoen, Benjie Chen, and David Mazi\`{e}res.
\newblock A low-bandwidth network file system.
\newblock In {\em Proceedings of the Eighteenth ACM Symposium on Operating
  Systems Principles}, SOSP '01, page 174–187, New York, NY, USA, 2001.
  Association for Computing Machinery.
\newblock \href {https://doi.org/10.1145/502034.502052}
  {\path{doi:10.1145/502034.502052}}.

\bibitem{nguyen2019}
Thuy~Linh Nguyen, Ramon Nou, and Adrien Lebre.
\newblock Yolo: Speeding up vm and docker boot time by reducing i/o operations.
\newblock In {\em European Conference on Parallel Processing}, pages 273--287.
  Springer, 2019.

\bibitem{virtio2016}
OASIS.
\newblock Virtual i/o device (virtio) version 1.0, March 2016.

\bibitem{Olson2021}
Marc Olson and Prarthana Karmakar.
\newblock Amazon ebs under the hood: A tech deep dive, December 2021.
\newblock URL: \url{https://www.youtube.com/watch?v=kaWzAEVZ6k8}.

\bibitem{ONeil1993}
Elizabeth~J O'neil, Patrick~E O'neil, and Gerhard Weikum.
\newblock The lru-k page replacement algorithm for database disk buffering.
\newblock {\em Acm Sigmod Record}, 22(2):297--306, 1993.

\bibitem{Quinlan2002}
Sean Quinlan and Sean Dorward.
\newblock Venti: A new approach to archival data storage.
\newblock In {\em Conference on File and Storage Technologies (FAST 02)},
  Monterey, CA, January 2002. USENIX Association.
\newblock URL:
  \url{https://www.usenix.org/conference/fast-02/venti-new-approach-archival-data-storage}.

\bibitem{Rashmi2016}
K.~V. Rashmi, Mosharaf Chowdhury, Jack Kosaian, Ion Stoica, and Kannan
  Ramchandran.
\newblock {EC-Cache}: {Load-Balanced}, {Low-Latency} cluster caching with
  online erasure coding.
\newblock In {\em 12th USENIX Symposium on Operating Systems Design and
  Implementation (OSDI 16)}, pages 401--417, Savannah, GA, November 2016.
  USENIX Association.
\newblock URL:
  \url{https://www.usenix.org/conference/osdi16/technical-sessions/presentation/rashmi}.

\bibitem{Russell2008}
Rusty Russell.
\newblock Virtio: Towards a de-facto standard for virtual i/o devices.
\newblock {\em SIGOPS Oper. Syst. Rev.}, 42(5):95--103, July 2008.
\newblock URL: \url{http://doi.acm.org/10.1145/1400097.1400108}, \href
  {https://doi.org/10.1145/1400097.1400108}
  {\path{doi:10.1145/1400097.1400108}}.

\bibitem{Storer2008}
Mark~W. Storer, Kevin Greenan, Darrell~D.E. Long, and Ethan~L. Miller.
\newblock Secure data deduplication.
\newblock In {\em Proceedings of the 4th ACM International Workshop on Storage
  Security and Survivability}, StorageSS '08, page 1–10, New York, NY, USA,
  2008. Association for Computing Machinery.
\newblock \href {https://doi.org/10.1145/1456469.1456471}
  {\path{doi:10.1145/1456469.1456471}}.

\bibitem{Thalheim2018}
J{\"o}rg Thalheim, Pramod Bhatotia, Pedro Fonseca, and Baris Kasikci.
\newblock Cntr: Lightweight {OS} containers.
\newblock In {\em 2018 USENIX Annual Technical Conference (USENIX ATC 18)},
  pages 199--212, Boston, MA, July 2018. USENIX Association.
\newblock URL:
  \url{https://www.usenix.org/conference/atc18/presentation/thalheim}.

\bibitem{Tokunaga2020}
Kohei Tokunaga.
\newblock Startup containers in lightning speed with lazy image distribution on
  containerd.
\newblock Accessed: 2022-04-15.
\newblock URL:
  \url{https://medium.com/nttlabs/startup-containers-in-lightning-speed-with-lazy-image-distribution-on-containerd-243d94522361}.

\bibitem{Vangoor2017}
Bharath Kumar~Reddy Vangoor, Vasily Tarasov, and Erez Zadok.
\newblock To {FUSE} or not to {FUSE}: Performance of {User-Space} file systems.
\newblock In {\em 15th USENIX Conference on File and Storage Technologies (FAST
  17)}, pages 59--72, Santa Clara, CA, February 2017. USENIX Association.
\newblock URL:
  \url{https://www.usenix.org/conference/fast17/technical-sessions/presentation/vangoor}.

\bibitem{Ashish2013}
Ashish Vulimiri, Philip~Brighten Godfrey, Radhika Mittal, Justine Sherry,
  Sylvia Ratnasamy, and Scott Shenker.
\newblock Low latency via redundancy.
\newblock In {\em Proceedings of the Ninth ACM Conference on Emerging
  Networking Experiments and Technologies}, CoNEXT '13, page 283–294, New
  York, NY, USA, 2013. Association for Computing Machinery.
\newblock \href {https://doi.org/10.1145/2535372.2535392}
  {\path{doi:10.1145/2535372.2535392}}.

\bibitem{Wang2021}
Ao~Wang, Shuai Chang, Huangshi Tian, Hongqi Wang, Haoran Yang, Huiba Li, Rui
  Du, and Yue Cheng.
\newblock {FaaSNet}: Scalable and fast provisioning of custom serverless
  container runtimes at alibaba cloud function compute.
\newblock In {\em 2021 USENIX Annual Technical Conference (USENIX ATC 21)},
  pages 443--457. USENIX Association, July 2021.
\newblock URL:
  \url{https://www.usenix.org/conference/atc21/presentation/wang-ao}.

\bibitem{Wu2015}
Zhe Wu, Curtis Yu, and Harsha~V. Madhyastha.
\newblock {CosTLO}: {Cost-Effective} redundancy for lower latency variance on
  cloud storage services.
\newblock In {\em 12th USENIX Symposium on Networked Systems Design and
  Implementation (NSDI 15)}, pages 543--557, Oakland, CA, May 2015. USENIX
  Association.
\newblock URL:
  \url{https://www.usenix.org/conference/nsdi15/technical-sessions/presentation/wu}.

\bibitem{Yang2020}
Juncheng Yang, Yao Yue, and K.~V. Rashmi.
\newblock A large scale analysis of hundreds of in-memory cache clusters at
  twitter.
\newblock In {\em 14th USENIX Symposium on Operating Systems Design and
  Implementation (OSDI 20)}, pages 191--208. USENIX Association, November 2020.
\newblock URL:
  \url{https://www.usenix.org/conference/osdi20/presentation/yang}.

\bibitem{Chao2018}
Chao Zheng, Lukas Rupprecht, Vasily Tarasov, Douglas Thain, Mohamed Mohamed,
  Dimitrios Skourtis, Amit~S. Warke, and Dean Hildebrand.
\newblock Wharf: Sharing docker images in a distributed file system.
\newblock In {\em Proceedings of the ACM Symposium on Cloud Computing}, SoCC
  '18, page 174–185, New York, NY, USA, 2018. Association for Computing
  Machinery.
\newblock \href {https://doi.org/10.1145/3267809.3267836}
  {\path{doi:10.1145/3267809.3267836}}.

\end{thebibliography}
\end{document}